\documentclass[a4paper,comsoc]{IEEEtran}
\IEEEoverridecommandlockouts
\usepackage{tikz}
\usepackage{pgfplots}
\usepackage{bm}
\usepackage{amsmath} 
\usepackage{amsthm}
\usepackage{amsfonts}
\usepackage{graphicx}
\usepackage{siunitx}
\usepackage{color}
\usepackage{fancyhdr}
\usepackage{array,multirow}
\usepackage{adjustbox}
\usepackage{balance}
\usepackage{cite}
\usepackage[acronym,shortcuts]{glossaries}
\usepackage{xspace}
\usepackage{tumcolor}
\usetikzlibrary{calc, shapes.geometric}

\fancypagestyle{cfooter}{ %
\fancyhf{} 
\cfoot{\scriptsize{\copyright\ 2024 IEEE. Personal use of this material is permitted. Permission from IEEE must be obtained for all other uses, in any current or future media, including reprinting / republishing this material for advertising or promotional purposes, creating new collective works, for resale or redistribution to servers or lists, or reuse of any copyrighted component of this work in other works.}}

}

\pgfplotsset{
	discard if not/.style 2 args={
		x filter/.code={
			\edef\tempa{\thisrow{#1}}
			\edef\tempb{#2}
			\ifx\tempa\tempb
			\else
			\fi
		}
	}
}
\pgfplotsset{compat=1.15}
\definecolor{mylila}{RGB}{153,50,204} 
\definecolor{mygreen}{RGB}{176,191,26} 

\def\pltw{250pt}
\def\plth{160pt}
\def\plthsmall{100pt}
\def\markSize{1.8pt}
\def\lineWidth{1.2pt}

\tikzset{VAEgenie/.style={mark options=solid, color=TUMBeamerRed, line width=\lineWidth, mark=Mercedes star, mark size=\markSize, solid}}
\tikzset{VAEnoisy/.style={mark options={solid}, color=TUMBeamerOrange, line width=\lineWidth, mark=triangle, mark size=\markSize, solid}}
\tikzset{VAEreal/.style={mark options={solid}, color=TUMBlue, line width=\lineWidth, mark=pentagon, mark size=\markSize, solid}}
\tikzset{VAErealvar/.style={mark options={solid}, color=mylila, line width=\lineWidth, mark=Mercedes star flipped, mark size=\markSize, solid}}
\tikzset{VAErealfix/.style={mark options={solid}, color=TUMGray, line width=\lineWidth, mark=square, mark size=\markSize, solid}}
\tikzset{geniecov/.style={mark options={solid}, color=blue, line width=\lineWidth, mark=x, mark size=\markSize, dashed}}
\tikzset{globalcov/.style={mark options={solid}, color=TUMMediumGray, line width=\lineWidth, mark=|, mark size=\markSize, dashed}}
\tikzset{LS/.style={mark options={solid}, color=black, line width=\lineWidth, mark=Mercedes star flipped, mark size=\markSize, dashed}}
\tikzset{GMM/.style={mark options={solid}, color=brown, line width=\lineWidth, mark=o, mark size=\markSize, dashed}}
\tikzset{{GMM circ}/.style={mark options={solid}, color=mylila, line width=\lineWidth, mark=diamond, mark size=\markSize, dashed}}
\tikzset{{GMM kron}/.style={mark options={solid}, color=brown, line width=\lineWidth, mark=o, mark size=\markSize, dashed}}
\tikzset{{GMM bcirc}/.style={mark options={solid}, color=mylila, line width=\lineWidth, mark=diamond, mark size=\markSize, dashed}}
\tikzset{CNN/.style={mark options={solid}, color=mylila, line width=\lineWidth, mark=10-pointed star, mark size=\markSize, dashed}}
\tikzset{AMP/.style={mark options={solid}, color=TUMBeamerLightBlue, line width=\lineWidth, mark=triangle, mark size=\markSize, dashed}}
\tikzset{{genie OMP}/.style={mark options={solid}, color=TUMBeamerDarkRed, line width=\lineWidth, mark=star, mark size=\markSize, dashed}}
\tikzset{{globalcovlin}/.style={mark options={solid}, color=brown, line width=\lineWidth, mark=o, mark size=\markSize, dashed}}
\tikzset{{linint}/.style={mark options={solid}, color=TUMBeamerLightBlue, line width=\lineWidth, mark=10-pointed star, mark size=\markSize, dashed}}

\newcommand{\legglobalcov}{LMMSE}
\newcommand{\legLS}{LS}

\newcommand{\legcnn}{CNN}

\newcommand{\leggenieomp}{genie-OMP}

\newcommand{\quadriga}{QuaDRiGa\xspace}

\newcommand{\Nv}{{N_\text{v}}}
\newcommand{\Nh}{{N_\text{h}}}
\newcommand{\NL}{{N_\text{L}}}
\newcommand{\Tr}{{T_\text{r}}}
\newcommand{\Tv}{{T_\text{v}}}
\newcommand{\Te}{{T_\text{e}}}
\newcommand{\cnd}{{\,|\,}}

\newcommand{\vmu}{{\bm{\mu}}}
\newcommand{\vsig}{{\bm{\sigma}}}

\newcommand{\vtheta}{{\bm{\theta}}}
\newcommand{\vphi}{{\bm{\phi}}}

\newcommand{\vc}{{\bm{c}}}

\newcommand{\vg}{{\bm{g}}}
\newcommand{\vh}{{\bm{h}}}

\newcommand{\vn}{{\bm{n}}}

\newcommand{\vy}{{\bm{y}}}
\newcommand{\vz}{{\bm{z}}}

\newcommand{\ma}{{\bm{A}}}

\newcommand{\mc}{{\bm{C}}}

\newcommand{\mq}{{\bm{Q}}}

\newcommand{\RR}{{\mathbb{R}}}
\newcommand{\CC}{{\mathbb{C}}}

\newcommand{\jim}{{\mathrm{j}}}
\newcommand{\OO}{{\mathcal O}}
\newcommand{\herm}{{^{\mkern0.5mu\mathrm{H}}}}
\newcommand{\tran}{{^{\mkern0.5mu\mathrm{T}}}}

\DeclareMathOperator{\E}{E}

\DeclareMathOperator{\diag}{diag}
\DeclareMathOperator{\KL}{D_{KL}}
\DeclareMathOperator*{\argmin}{arg\,min}

\newacronym{tdd}{TDD}{time division duplex}
\newacronym{fdd}{FDD}{frequency division duplex}
\newacronym{lmmse}{LMMSE}{linear minimum mean squared error}
\newacronym{mse}{MSE}{mean squared error}
\newacronym{mmse}{MMSE}{minimum mean squared error}
\newacronym{nmse}{NMSE}{normalized mean squared error}
\newacronym{mimo}{MIMO}{multiple-input multiple-output}
\newacronym{simo}{SIMO}{single-input multiple-output}
\newacronym{miso}{MISO}{multiple-input single-output}
\newacronym{siso}{SISO}{single-input single-output}
\newacronym{deep}{DL}{deep learning}
\newacronym{ofdm}{OFDM}{orthogonal frequency division multiplexing}
\newacronym{csi}{CSI}{channel state information}
\newacronym{ula}{ULA}{uniform linear array}
\newacronym{ura}{URA}{uniform rectangular array}
\newacronym{dft}{DFT}{discrete fourier transform}
\newacronym{bs}{BS}{base station}
\newacronym{mt}{MT}{mobile terminal}
\newacronym{ae}{AE}{autoencoder}
\newacronym{ml}{ML}{machine learning}
\newacronym{dl}{DL}{deep learning}
\newacronym{doa}{DoA}{direction of arrival}
\newacronym{dod}{DoD}{direction of departure}
\newacronym{kl}{KL}{Kullback-Leibler}
\newacronym{elbo}{ELBO}{evidence-lower bound}
\newacronym{iid}{i.i.d.}{independent and identically distributed}
\newacronym{fc}{FC}{fully connected}
\newacronym{nn}{NN}{neural network}
\newacronym{dnn}{DNN}{deep neural network}
\newacronym{cnn}{CNN}{convolutional neural network}
\newacronym{ls}{LS}{least squares}
\newacronym{snr}{SNR}{signal-to-noise ratio}
\newacronym{ce}{CE}{channel estimation}
\newacronym{ul}{UL}{uplink}
\newacronym{mpc}{MPC}{multipath component}
\newacronym{rt}{RT}{ray-tracing}
\newacronym{mmd}{MMD}{maximum mean discrepancy}
\newacronym{cdf}{CDF}{cumulative distribution function}
\newacronym{tpr}{TPR}{true positive rate}
\newacronym{ccm}{CCM}{channel covariance matrix}
\newacronym{cg}{CG}{conditionally Gaussian}
\newacronym{3gpp}{3GPP}{3rd Generation Partnership Project}
\newacronym{vae}{VAE}{variational autoencoder}
\newacronym{vi}{VI}{variational inference}
\newacronym{cc}{CC}{convolutional channel}
\newacronym{cl}{CL}{convolutional layer}
\newacronym{ll}{LL}{linear layer}
\newacronym{rl}{RL}{reshaping layer}
\newacronym{bn}{BN}{batch normalization}
\newacronym{cs}{CS}{compressed sensing}
\newacronym{amp}{AMP}{approximate message passing}
\newacronym{gmm}{GMM}{Gaussian mixture model}
\newacronym{nf}{NF}{normalizing flow}
\newacronym{gan}{GAN}{generative adversarial network}
\newacronym{vdm}{VDM}{variational diffusion model}
\newacronym{cme}{CME}{conditional mean estimator}
\newacronym{los}{LOS}{line of sight}
\newacronym{nlos}{NLOS}{non-line of sight}
\newacronym{omp}{OMP}{orthogonal matching pursuit}
\newacronym{li}{LI}{linear interpolation}
\newacronym{mb}{MB}{model-based}
\newacronym{ud}{UD}{underdetermined}
\newacronym{fd}{FD}{fully-determined}
\newacronym{umi}{UMi}{Urban Mircocell}
\newacronym{uma}{UMa}{Urban Macrocell}
\newacronym{awgn}{AWGN}{additive white Gaussian noise}
\def\BibTeX{{\rm B\kern-.05em{\sc i\kern-.025em b}\kern-.08em
    T\kern-.1667em\lower.7ex\hbox{E}\kern-.125emX}}

\begin{document}

\title{Variational Autoencoder for Channel Estimation: Real-World Measurement Insights
\thanks{This work is supported by the Bavarian Ministry of Economic Affairs, Regional Development and Energy within the project 6G Future Lab Bavaria. The authors acknowledge the financial support by the Federal Ministry of Education and Research of Germany in the program of “Souverän. Digital. Vernetzt.”. Joint project 6G-life, project identification number: 16KISK002}
}

\author{\IEEEauthorblockN{Michael Baur, Benedikt Böck, Nurettin Turan, and Wolfgang Utschick}\\
\IEEEauthorblockA{TUM School of Computation, Information and Technology, Technical University of Munich, Germany}\\
Email: \{mi.baur, benedikt.boeck, nurettin.turan, utschick\}@tum.de
}

\maketitle
\thispagestyle{cfooter}
\begin{abstract}
This work utilizes a variational autoencoder for channel estimation and evaluates it on real-world measurements. The estimator is trained solely on noisy channel observations and parameterizes an approximation to the mean squared error-optimal estimator by learning observation-dependent conditional first and second moments. The proposed estimator significantly outperforms related state-of-the-art estimators on real-world measurements. We investigate the effect of pre-training with synthetic data and find that the proposed estimator exhibits comparable results to the related estimators if trained on synthetic data and evaluated on the measurement data. Furthermore, pre-training on synthetic data also helps to reduce the required measurement training dataset size.
\end{abstract}

\section{Introduction}
\label{sec:intro}

\Ac{mb} \ac{dl} is an intriguing paradigm for the design of novel algorithms~\cite{Shlezinger2023}, which stands in opposition to end-to-end learning~\cite{Yu2022}.
Instead of designing a task-specific \ac{dnn} whose aim is to learn most functionalities from scratch as most end-to-end approaches do, \ac{mb}-\ac{dl} utilizes established relationships and replaces only specific parts within the processing chain with a \ac{dnn}.
For instance, in \ac{ce}, structural knowledge about a suboptimal approximation to the \ac{mmse} estimator is used for the blueprint of a \ac{cnn}~\cite{Neumann2018} or a \ac{dnn} serves as a denoiser for the LDAMP algorithm~\cite{He2018a}.

The recently proposed \ac{vae}-based channel estimator is a \ac{mb}-\ac{dl} method for \ac{ce}~\cite{Baur2022,Baur2023}, which can also be applied to estimation problems in time-varying and frequency-selective situations~\cite{Bock2023,Baur2023a}.
The concept is to train a \ac{vae} on \ac{csi} data stemming from and representing a radio propagation environment to learn the underlying distribution.
A remarkable property of the \ac{vae} is that it can be trained solely with noisy channel observations without access to perfect \ac{csi} during the training.
Subsequently, the trained \ac{vae} is used to provide input data dependent conditional first and second moments to parameterize an approximation to the \ac{mse}-optimal \ac{cme}.
The \ac{vae} is a so-called generative model and was applied to a variety of other communications-related problems, e.g., channel equalization~\cite{Lauinger2022a} or channel modeling~\cite{Xia2022a}.
Another prominent generative model representative is the \ac{gan}~\cite{Goodfellow2020}.
Particularly, the \ac{gan} is applied to \ac{ce} in hybrid, wideband, and quantized systems~\cite{Balevi2021,Doshi2022,Doshi2023}.

The \ac{vae}-based channel estimator from~\cite{Baur2023} demonstrates excellent \ac{ce} results for realistic but, so far, synthetic channel data.
Therefore, this work evaluates the \ac{vae}-based channel estimator on real-world measurement data, which was previously used to assess other data-driven estimators~\cite{Hellings2019,Turan2022}. 
The transition from synthetic to measured data is essential because many idealistic modeling assumptions in the synthetic case do not hold in reality.
An example of such an idealistic assumption is array imperfections. 
We attain the following main insights from the evaluation with the measured data:
\begin{itemize}
    \item The \ac{vae}-based estimator significantly outperforms the related estimators.
    \item For the best estimation quality, a large training dataset is necessary.
    \item A \ac{vae} trained on synthetic and evaluated on measured data shows comparable results as the best-performing related estimators.
    \item Pre-training with synthetic training data notably lowers the required measurement training dataset size.
\end{itemize}

\section{System Model and Problem Formulation}
\label{sec:system}

We consider the uplink of a \ac{simo} communications system with block-fading.
The \ac{bs} is equipped with a \ac{ura} and has $\Nv$ antennas in its vertical and $\Nh$ antennas in its horizontal direction.
Let $N=\Nv\Nh$.
The \ac{mt} is equipped with a single antenna.
Under the assumption of a single-user scenario without pilot contamination, the \ac{bs} receives a noisy pilot signal from the \ac{mt} in the far field, which reads as
\begin{equation}
    \vy = \vh + \vn, \quad \vy\in\CC^{N},
    \label{eq:system}
\end{equation}
after decorrelating the single pilot. 
The frequency-flat assumed channel $\vh$ follows the unknown prior $p(\vh)$ and is perturbed by \ac{awgn} $\vn\sim\mathcal{N}_\CC(\bm{0},\varsigma^2\mathbf I)$.

In the far field, the \ac{ccm} of the channel entries in either vertical or horizontal direction is Toeplitz due to the uniform spacing at the \ac{ura}.
Consequently, the \ac{ccm} of all channel entries is block-Toeplitz. 
A block-Toeplitz structure can be enforced with the parameterization~\cite{Strang1986}: 
\begin{equation}
    \mq\herm \diag(\vc) \mq, \quad \vc\in\RR^{4N}_+.
    \label{eq:btoep}
\end{equation}
The matrix $\mq=\mq_{\Nv}\otimes\mq_{\Nh}$, $\mq_{\Nv}\in\CC^{2\Nv\times\Nv}$ contains the first $\Nv$ columns of the $2\Nv\times2\Nv$ \ac{dft} matrix, and $\mq_{\Nh}$ is defined accordingly. 

In \ac{ce}, the goal is to estimate $\vh$ based on $\vy$ from~\eqref{eq:system}.
The \ac{cme} 
\begin{equation}
    \hat{\vh}_{\text{\normalfont CME}}(\vy) = \E[\vh\cnd\vy] = \argmin_{\hat\vh\in\CC^N}\E\left[ \| \vh - \hat\vh \|^2 \right]
    \label{eq:cme-mse}
\end{equation}
results in \ac{mse}-optimal channel estimates~\cite[Ch.~11]{Kay1993}.
By applying Bayes rule to $p(\vh\cnd\vy)$ outcomes
\begin{equation}
    p(\vh\cnd\vy) = \frac{p(\vy\cnd\vh)\, p(\vh)}{p(\vy)} = \frac{p_\vn(\vy - \ma\vh)\, p(\vh)}{p(\vy)}, 
\end{equation}
which we utilize to reformulate the \ac{cme} as
\begin{equation}
    \E[\vh\cnd\vy] = \int \vh \frac{p_\vn(\vy - \ma\vh)\, p(\vh)}{p(\vy)} \mathrm{d} \vh.
    \label{eq:cme-bayes}
\end{equation}
Note that $p_\vn$ represents the distribution of $\vn$.
Eq.~\eqref{eq:cme-bayes} reveals why the \ac{cme} is hard to obtain in practice.
It needs access to $p(\vh)$, i.e., $p(\vh)$ must be estimated directly or indirectly.
Furthermore, the integral in~\eqref{eq:cme-bayes} needs to be approximated since, in general, there exists no closed-form solution.
The \ac{vae} is a generative model whose objective is the approximation of $p(\vh)$, and it provides a practical way to locally approximate the integral in~\eqref{eq:cme-bayes}.
These properties motivate the application of the \ac{vae} to \ac{ce}.

\section{VAE-based Channel Estimation}
\label{sec:method}

\subsection{Preliminaries}
\label{subsec:prelim}

The elaborations in this and the following section follow the description of the \ac{vae}-based channel estimator in~\cite{Baur2023,Baur2023a}.
The \ac{elbo} is the central term for the training of a \ac{vae} and is a lower bound to a parameterized likelihood model $p_\vtheta(\vy)$ of the unknown distribution $p(\vy)$.
An accessible version of the \ac{elbo} reads as~\cite{Kingma2019}
\begin{equation}
    \mathcal{L}_{\vtheta,\vphi}(\vy) = \E_{q_{\vphi}} \left[ \log p_{\vtheta}(\vy\cnd\vz) \right] - \KL(q_{\vphi}(\vz\cnd\vy)\,\|\,p(\vz))
    \label{eq:vae}
\end{equation}
with $\E_{q_\vphi(\vz|\vy)}[\cdot] = \E_{q_\vphi}[\cdot]$ as the expectation according to the variational distribution $q_\vphi(\vz\cnd\vy)$, which is supposed to approximate $p_\vtheta(\vz\cnd\vy)$.
Note that the parameterization of $p_\vtheta(\vz\cnd\vy)$ is determined by the joint $p_\vtheta(\vy,\vz)$ whose parameterizaiton is set by $p_\vtheta(\vy\cnd\vz)$.
The latent vector $\vz\in\RR^\NL$ is introduced in the \ac{vae} to tractably improve the expressiveness of the parameterized likelihood model.
The last term in~\eqref{eq:vae} is the \ac{kl} divergence
\begin{equation}
    \KL(q_\vphi(\vz\cnd\vy)\,\|\,p_\vtheta(\vz)) = \E_{q_\vphi}\left[ \log \left( \frac{q_\vphi(\vz\cnd\vy)}{p_\vtheta(\vz)} \right) \right].
    \label{eq:elbo-gap}
\end{equation}

\begin{figure}[t]
  \centering
  \begin{tikzpicture}[>=stealth, scale=0.9]
	\def\NNsize{2cm}
	\def\NNwidth{2cm}
	\def\NNheight{2.2cm}
	\def\NNangle{70}
	\node (input) at (0, 0) { $\bm{y}$};
	
	\node[trapezium, draw, align=center, trapezium stretches = true, minimum height=\NNwidth, minimum width=\NNheight, align=center, trapezium angle=\NNangle, rotate=-90] (NN1) at($ (input) + (2,0) $) {\rotatebox{90}{Encoder} \small{ \rotatebox{90}{\hspace{-1mm} $q_{\vphi}(\bm{z}\cnd\bm{y})$}}};
	
	\node[circle, draw, inner sep = 0.01em] (add) at ($ (NN1.north) + (1.3,0.4) $) {$ + $};
	\node[circle, draw, inner sep = 0.01em] (multi) at ($ (NN1.north) + (1.3,-0.4) $) {$ \odot $};
	\node (eps) at ( $(multi) + (0.3, -1)$ ) {\hspace{8.5mm} {$\bm{\varepsilon} \sim \mathcal{N}(\bm{0},\mathbf{I}) $}};
	
	\node[trapezium, draw, align=center, trapezium stretches = true, minimum height=\NNwidth, minimum width=\NNheight, align=center, trapezium angle=\NNangle, rotate=90](NN2) at ($ (NN1) + (4.6,0) $) {\small{\rotatebox{-90}{\hspace{0.25mm}$p_{\vtheta}(\bm{y}\cnd\bm{z})$}}\hspace{1mm} \normalsize{\rotatebox{-90}{Decoder}}};
	
	
	\node(output_mu) at ($ (NN2.south) + (1.3,0.4) $) { $\vmu_\vtheta(\vz)$};
	\node(output_cov) at ($ (NN2.south) + (1.3,-0.4) $) { $\mc_\vtheta(\vz)$};
	
	\draw[->] (input.east) -- (NN1.south) {};
	
	\draw[->] ($ (NN1.north) + (0,0.4) $) --node[midway, above=-.0em]{ $\vmu_\vphi(\vy)$}    (add.west);
	\draw[->] ($ (NN1.north) + (0,-0.4) $) --node[midway, above=-1.9em]{ $\vsig_\vphi(\vy)$} (multi.west);
	\draw[->] (multi.north) -- (add.south);
	\draw[->] ($ (eps.north) - (0.3, 0.1) $)   -- (multi.south);
	\draw[->] (add.east) --node[midway, above=.1em]{ $\bm{z}$} ($(NN2.north) + (0,0.4)$);
	
	\draw[->] ($ (NN2.south) + (0,0.4) $) -- (output_mu.west);
	\draw[->] ($ (NN2.south) + (0,-0.4) $) -- (output_cov.west);
\end{tikzpicture}
\vspace{-5mm}
\caption{Structure of a VAE with CG distributions $q_{\vphi}(\vz\cnd\vy)$ and $p_{\vtheta}(\vy\cnd\vz)$. The encoder and decoder represent \acp{dnn}.}
\label{fig:vae}
\end{figure}
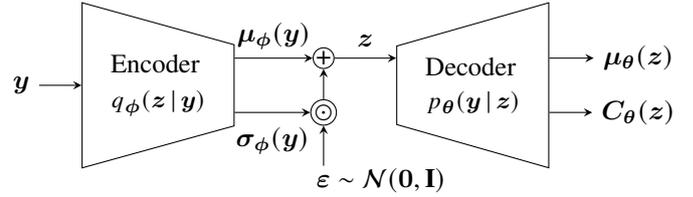

The \ac{vae} optimizes the \ac{elbo} with the help of \acp{dnn} and the reparameterization trick~\cite{Kingma2014}.
To this end, it is necessary to define the involved distributions, which we fulfill as follows: 
\begin{align}
    p(\vz) =\ & \mathcal{N}(\bm{0},\mathbf{I}), \nonumber \\
    p_{\vtheta}(\vy\cnd\vz) =\ & \mathcal{N}_{\CC}(\vmu_\vtheta(\vz),\tilde\mc_\vtheta(\vz)), \label{eq:dist_vae} \\  
    q_{\vphi}(\vz\cnd\vy) =\ & \mathcal{N}(\vmu_\vphi(\vy),\diag(\vsig^2_\vphi(\vy))), \nonumber 
\end{align}
and $\tilde\mc_\vtheta(\vz) = \mc_\vtheta(\vz) + \varsigma^2\mathbf{I}$.
Both $p_{\vtheta}(\vy\cnd\vz)$ and $q_{\vphi}(\vz\cnd\vy)$ are therefore \ac{cg} distributions, realized by \acp{dnn} with parameters $\vtheta$ and $\vphi$.
Accordingly, we obtain closed-form expressions for the terms in~\eqref{eq:vae}, i.e., $\left( -\E_{q_{\vphi}} \left[\log p_{\vtheta}(\vy\cnd\vz)\right] \right)$ is replaced by the estimate
\begin{equation}
     \log\det(\pi\,\tilde\mc_\vtheta(\tilde\vz)) + (\vh - \vmu_\vtheta(\tilde\vz))\herm \tilde\mc^{-1}_\vtheta(\tilde\vz) (\vh - \vmu_\vtheta(\tilde\vz))
    \label{eq:vae-dec-like}
\end{equation}
with the single monte carlo sample $\tilde\vz\sim q_{\vphi}(\vz\cnd\vy)$. 
The term $\KL(q_\vphi(\vz\cnd\vy)\,\|\,p_\vtheta(\vz))$ is the \ac{kl} divergence between two Gaussian distributions, which becomes
\begin{equation}
    \frac{1}{2} \left( \bm{1}\tran \left( -\log\vsig^2_\vphi(\vy) + \vmu_\vphi(\vy)^2 + \vsig^2_\vphi(\vy) \right) - \NL \right),
    \label{eq:vae-kl}
\end{equation}
where $\bm 1$ represents the all-ones vector.
A schematic illustration of a \ac{vae} implementation is displayed in Fig.~\ref{fig:vae}. 
The presented \ac{vae} consists of an encoder that represents $q_\vphi(\vz\cnd\vy)$ and outputs $\{\vmu_\vphi(\vy),\vsig_\vphi(\vy)\}$ and a decoder that represents $p_{\vtheta}(\vy\cnd\vz)$ and outputs $\{\vmu_\vtheta(\vz),\mc_\vtheta(\vz)\}$.
The encoder receives $\vy$ as input and produces with its outputs the reparameterized sample $\vz$, which is the decoder input.
We refer the reader to~\cite{Baur2023,Kingma2019} for a more detailed introduction to the \ac{vae} framework.

\subsection{MMSE Estimation with the VAE}
\label{subsec:mmse-est-vae}

After successfully training the \ac{vae}, we obtain a generative model that locally parameterizes $p(\vy)$ as \ac{cg}. 
For estimation purposes, it is indeed more desirable to have a parametric model for $p(\vh)$.
We can achieve this for the system model in~\eqref{eq:system} by requiring the \ac{vae} decoder outputs $\vmu_\vtheta(\vz)$ and $\mc_\vtheta(\vz)$ to be the mean and covariance of $\vh\cnd\vz$ instead of $\vy\cnd\vz$.
Since we have an \ac{awgn} channel, where knowledge of the noise variance is assumed, we add $\varsigma^2\mathbf{I}$ to $\mc_\vtheta(\vz)$ for the computation of~\eqref{eq:vae-dec-like} during the training to acquire the intended outcome.
It follows that the \ac{vae}'s goal is to provide
\begin{equation}
    \vh \mid \vz \sim p_\vtheta(\vh\cnd\vz) = \mathcal{N}_{\CC}(\vmu_\vtheta(\vz), \mc_\vtheta(\vz))
    \label{eq:cg-vae}
\end{equation}
after the training.
The \ac{vae} learns abstract conditions $\vz$ in its latent space that may not necessarily carry a physical interpretation.
The advantage of the abstractness of the condition is the flexibility to model the channels as \ac{cg} since the model can select an almost arbitrary condition for every channel.
The \ac{cg} property will be central for the design of the channel estimator.

Section~\ref{sec:system} describes that the \ac{cme} delivers MMSE channel estimates. 
The \ac{cme} can be reformulated with the law of total expectation as
\begin{equation}
    \E[\vh\cnd\vy] = \E_{p_\vtheta(\vz\cnd\vy)}\left[ \E[\vh\cnd\vz,\vy]\cnd\vy \right]. 
    \label{eq:total_exp}
\end{equation}
Further, let $t_\vtheta( \vz, \vy) = \E[\vh\cnd\vz,\vy]$.
Assuming that~\eqref{eq:cg-vae} holds, a closed-form solution for $t_\vtheta( \vz, \vy)$ exists due to the \ac{cg} property.
Therefore~\cite[Ch.~11]{Kay1993},
\begin{equation}
    t_\vtheta( \vz, \vy) = \vmu_\vtheta(\vz) + \mc_\vtheta(\vz) (\mc_\vtheta(\vz) + \varsigma^2\mathbf{I})^{-1} (\vy - \vmu_\vtheta(\vz)),
    \label{eq:lmmse-vae}
\end{equation}
with $\vy$ from~\eqref{eq:system}.
The latent vector $\vz$ depends on $\vy$ through the encoder.
The quantities $\vmu_\vtheta(\vz)$ and $\mc_\vtheta(\vz)$ are provided by the decoder of the VAE.
It remains to compute the outer expectation in~\eqref{eq:total_exp}, with respect to $p_\vtheta(\vz\cnd\vy)$.
Since $p_\vtheta(\vz\cnd\vy)$ is approximated by $q_\vphi(\vz\cnd\vy)$, we can approximate the outer expectation with samples from $q_\vphi(\vz\cnd\vy)$.
As analyzed in detail in~\cite{Baur2023}, the single sample $\vmu_\vphi(\vz)$ from $q_\vphi(\vz\cnd\vy)$ achieves excellent \ac{ce} performance. 
With all the previous depictions, we obtain the \ac{vae}-based estimator
\begin{equation}
    \hat\vh_{\text{VAE}}(\vy) =  t_\vtheta( \vz=\vmu_\vphi(\vy), \vy).
    \label{eq:vae-estimator-mu}
\end{equation}

The estimator $\hat\vh_{\text{VAE}}(\vy)$ requires an offline training phase of the \ac{vae} in Fig.~\ref{fig:vae} to provide the necessary first and second conditional moment for the evaluation of $t_\vtheta( \vz, \vy)$.
An intriguing aspect of the estimator in this work is that its training is based solely on noisy observations $\vy$ and not on perfect \ac{csi} data $\vh$.
This contrasts the plethora of \ac{dl}-based estimators, that usually require perfect \ac{csi} during the training, which is only available after measurement campaigns.
We thus think that $\hat\vh_{\text{VAE}}(\vy)$ is a realistic estimator that matches real-world necessities.
The estimator $\hat\vh_{\text{VAE}}(\vy)$ may also be extended to underdetermined systems, e.g., for hybrid and wideband systems~\cite{Baur2023a}.
However, the treatment of such cases exceeds the scope of this work.
Previous work demonstrated that $\hat\vh_{\text{VAE}}(\vy)$ achieves excellent \ac{ce} performance for the \ac{simo} system model~\cite{Baur2022,Baur2023}, which we want to verify for real-world measurements in this work.

\subsection{Practical Considerations}
\label{subsec:practical}

For the evaluation of $\hat\vh_{\text{VAE}}(\vy)$, the matrix $\mc_\vtheta(\vz)$ must approximate the true \ac{ccm} as well as possible. 
Since we have a \ac{ura} at the \ac{bs}, we can exploit the resulting block-Toeplitz structure of the \ac{ccm} in~\eqref{eq:btoep} for the parameterization of $\mc_\vtheta(\vz)$.
Straightforwardly, we propose to approximate the \ac{ura} \ac{ccm} in the \ac{vae} framework with
\begin{equation}
    \mc_\vtheta(\vz) = \mq\herm \diag(\vc_\vtheta(\vz)) \mq, \quad \vc_\vtheta(\vz)\in\RR^{4N}_+,
    \label{eq:btoep-vae}
\end{equation}
where $\mq$ is defined as in~\eqref{eq:btoep}.
The vector $\vc_\vtheta(\vz)$ is outputted by the \ac{vae} decoder.

The computational complexity of an estimator is an essential aspect for \ac{ce}.
The complexity of $\hat\vh_{\text{VAE}}(\vy)$ consists of two parts: a forward pass of the \ac{vae} to yield the decoder outputs and one evaluation of~\eqref{eq:lmmse-vae}.
For the \ac{vae} forward pass, we can utilize the analysis from~\cite{Baur2023}, as the \ac{vae} architecture is conceptually the same.
The complexity for the $D$ layer \ac{vae} was found to be $\OO(DN^2)$, which is a consequence of the final \ac{ll} that yields the decoder outputs.
This outcome also holds for the model in this work.
The second part of the computational complexity relates to the evaluation of~\eqref{eq:lmmse-vae} and is dominated by the inversion of the matrix $\mc_\vtheta(\vz)+\varsigma^2\mathbf{I}$.
The inversion of this positive-definite block-Toeplitz matrix can be completed in $\OO(N^2)$ time with the Levinson algorithm~\cite[Ch.~6]{Kay1988}.
In conclusion, the worst-case complexity for the computation of $\hat\vh_{\text{VAE}}(\vy)$ is $\OO(DN^2)$, which can be further reduced by parallel computations and network pruning of the \ac{dnn}.

\section{CSI Data Aquisition}
\label{sec:channel}

\subsection{Measurement Campaign}
\label{subsec:campaign}

\begin{figure}[t]
    \centering
    \begin{tikzpicture}
        \draw (0,0) node[below right] {\includegraphics[width=0.975\columnwidth]{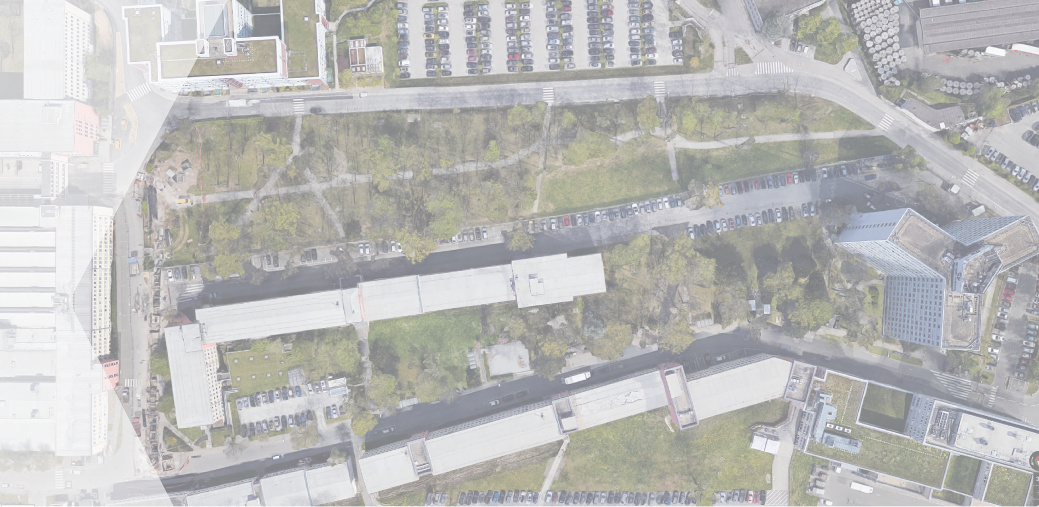}};
        \draw (0,0) node[below right] {\includegraphics[width=0.975\columnwidth]{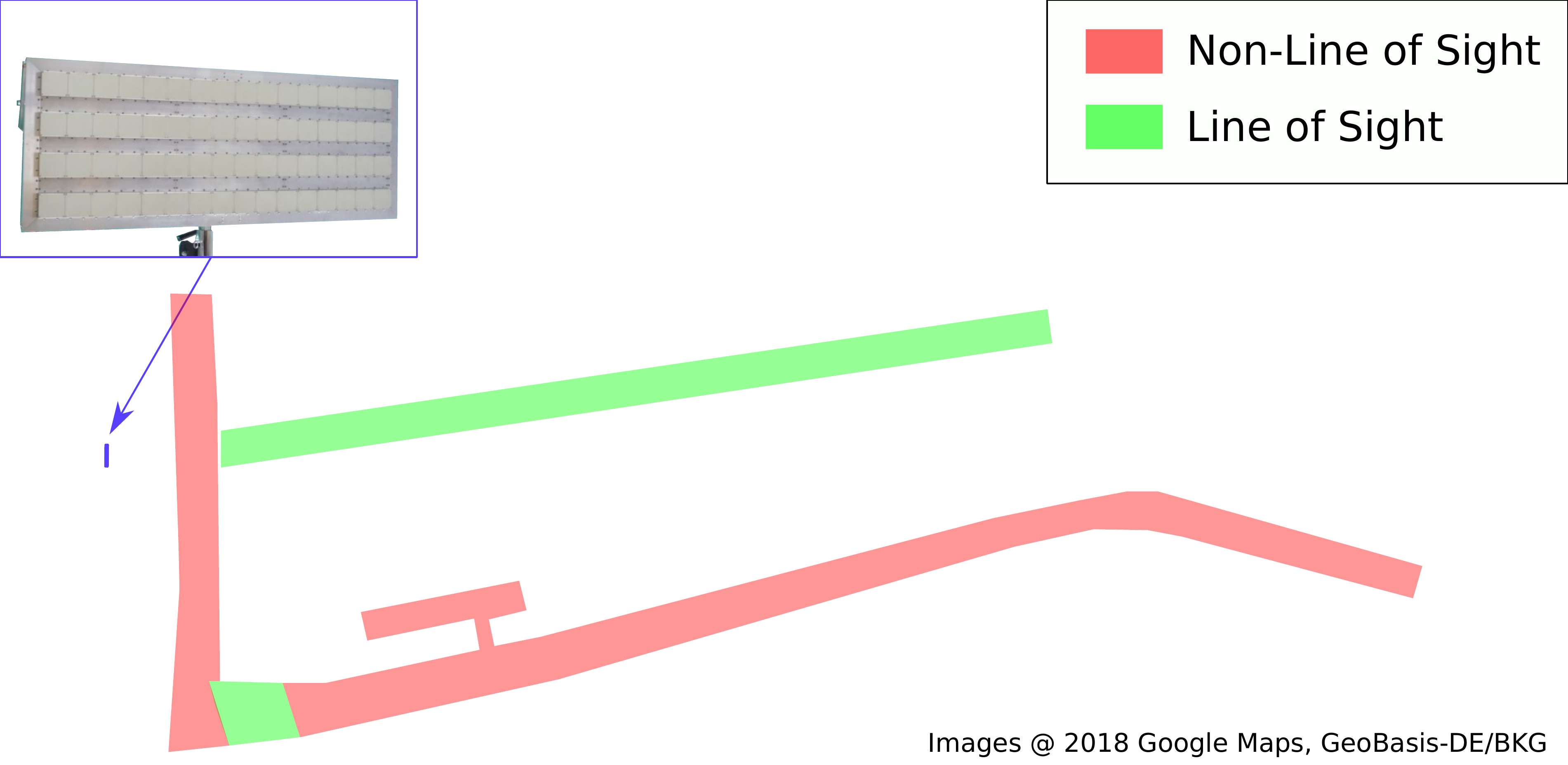}};
    \end{tikzpicture}
    \caption{Measurement setup on the Nokia campus in Stuttgart, Germany.}
    \label{fig:meas_campaign}
\end{figure}

As described in~\cite{Hellings2019,Turan2022}, the measurement campaign was conducted at the Nokia campus in Stuttgart, Germany, in 2017.
Fig.~\ref{fig:meas_campaign} displays the scenario.
On a rooftop roughly \SI{20}{m} above the ground, the \ac{bs} is located with a \SI{10}{\degree} down-tilt.
The \ac{bs} is a $4\times 16$ \ac{ura} with horizontal single polarized patch antennas and was adapted to match the \ac{3gpp} \ac{umi} propagation scenario.
As a result, $\Nv=4$ and $\Nh=16$.
The antenna spacing is $\lambda$ in the vertical and $\lambda/2$ in the horizontal direction, with $\lambda$ being the wavelength.
The single monopole receive antenna representing the \ac{mt} was placed on a moving vehicle with a maximum speed of \SI{25}{km\per h}.
GPS was used for the synchronization between the transmitter and receiver.
The data was collected by a TSMW receiver and stored on a Rohde \& Schwarz IQR hard disk recorder.
The carrier frequency was \SI{2.18}{\giga\hertz}.
The \ac{bs} transmitted \SI{10}{\mega\hertz} \ac{ofdm} waveforms with $600$ subcarriers in \SI{15}{\kilo\hertz} spacing.
The pilots were sent continuously with a periodicity of \SI{0.5}{ms}, arranged in $50$ separate subbands, with $12$ consecutive subcarriers each.
The propagation channel was assumed to remain constant for one pilot burst.

Channel realization vectors with $64$ coefficients per subband were extracted in a post-processing step.
The measurement campaign was conducted at a high \ac{snr} between $20$ and \SI{30}{dB}.
Such high \acp{snr} are necessary to obtain almost perfect \ac{csi} data needed for the offline training of most \ac{dl}-based channel estimators, e.g.~\cite{Ye2018,Soltani2019,Neumann2018}.
We corrupt the measured channels with \ac{awgn} at specific \ac{snr} values to generate noisy observations according to~\eqref{eq:system}.
For the training of the \ac{vae}-based channel estimator presented in Section~\ref{subsec:mmse-est-vae}, it is not required to actually conduct a measurement campaign at such high \ac{snr} values to obtain almost perfect \ac{csi} data.
Instead, it would suffice to gather a dataset of noisy observations $\vy$ from~\eqref{eq:system} at \ac{snr} values that are present in the online estimation phase received and collected at the \ac{bs} during regular operation.

\subsection{Synthetic Data}

Besides measured data, we also need synthetic \ac{csi} data for the analysis in Section~\ref{sec:results}.
To this end, we utilize the \quadriga channel simulator in version 2.6.1 to generate artificial channel vector realizations~\cite{Jaeckel2014}.
The channels are supposed to belong to a \ac{3gpp} \ac{umi} scenario, and the environment in \quadriga is designed to approximate the measurement site in Fig.~\ref{fig:meas_campaign} as close as possible.
This includes using the same carrier frequency of \SI{2.18}{\giga\hertz}, placing the \ac{bs} at a similar height and distance to the \acp{mt}, and assigning the \acp{mt} a comparable velocity.
The \ac{bs} is also equipped with a \ac{ura} and comprises identical dimensions and spacings as in the measurement data.
\quadriga models the channels as a superposition of in total $L$ propagation paths, where $L$ is determined whether \ac{los} or \ac{nlos} conditions are present.
More precisely, a channel is given as $\vh = \sum_{\ell=1}^{L} \vg_{\ell} \exp({-2\pi \jim f_c \tau_{\ell}})$, with $f_c$ as the carrier frequency and $\tau_\ell$ as the delay of the $\ell$-th propagation path.
The vector $ \vg_{\ell} $ accounts for the path attenuation between the \ac{mt} and every receive antenna, the antenna radiation pattern, and the polarization.
Eventually, the generated channels are post-processed to normalize the path gains.
\section{Simulation Results}
\label{sec:results}

\subsection{Implementation Details}
\label{subsec:implementation}

For the measurement data, we obtain a dataset of size $420{,}000$ of which we use $\Tr=400{,}000$ samples for training, $\Tv=10{,}000$ for validation, and $\Te=10{,}000$ for testing. 
For the synthetic \quadriga data, we generate in total $520{,}000$ channels, which are split into $500{,}000$, $10{,}000$, and $10{,}000$ training, validation, and test samples, respectively.
The channels are normalized in each dataset such that $\E[\|\vh\|^2]=N$. 
Accordingly, we define the \ac{snr} as $1/\varsigma^2$.
We evaluate our methods with the \ac{nmse} 
\begin{equation}
    \text{NMSE} = \frac{1}{T_{\text{e}}N} \sum_{i=1}^{T_{\text{e}}}\|\vh_i-\hat{\vh}_i\|^2,    
\end{equation}
where $\vh_i$ is the $i$-th channel realization in the test dataset, and $\hat{\vh}_i$ the corresponding estimate.

Conceptually, the \ac{vae} architecture is identical to the one presented in~\cite{Baur2023}.
The real and imaginary parts of $\vy$ are stacked as \acp{cc} and serve as \ac{vae} encoder input.
A $1\times 1$ \ac{cl} follows that maps to \textsc{ch} \acp{cc} to which we refer in Table~\ref{tab:parameters}.
Three building blocks consisting of a \ac{cl}, a \ac{bn}, and a ReLU activation function follow, where the \ac{cc} number is multiplied by $1.75$ and rounded after each \ac{cl}.
A subsequent \ac{ll} maps the flattened vector to the encoder outputs $\{\vmu_\vphi(\vy),\vsig_\vphi(\vy)\}$.
The decoder is a symmetrically flipped version of the encoder, which implies that all \acp{cl} are replaced with transposed \acp{cl}.
The remaining layer types remain unchanged.
The final transposed \ac{cl} maps to three \acp{cl}.
A flattening follows and a final \ac{ll} maps to $\{\vmu_\vtheta(\vz),\vc_\vtheta(\vz)\}$.
We enforce strictly positive values in $\vsig_\vphi(\vy)$ and $\vc_\vtheta(\vz)$ with an exponential function.
In all \acp{cl}, we use 1D \acp{cl} with a kernel size of eleven and a stride of two.
For the optimization, we use Adam as optimizer with a learning rate of $5\cdot10^{-4}$ and a batch size of 256.
The latent space dimensionality $\NL$ is $32$.
The implementation is realized with \textit{PyTorch}.
Unless stated otherwise, we train each \ac{vae} until it does not improve in terms of \ac{nmse} on the validation dataset for $100$ consecutive epochs.

\subsection{Architecture Study}
\label{subsec:architecture}

\begin{table}[t]
\centering
\caption{Performance of VAEs with increasing parameter amount.}
\renewcommand{\arraystretch}{1.3}
\label{tab:parameters}
\begin{tabular}{c|r|r|r|r}
\hline
\multicolumn{1}{c|}{VAE} & \multicolumn{1}{c|}{\textsc{ch}} & \multicolumn{1}{c|}{parameters} & \multicolumn{1}{c|}{NMSE at 10\,dB} & \multicolumn{1}{c}{NMSE at 20\,dB} \\ \hline
1 & $4$ & $319{,}291$ & $3.574\cdot 10^{-2}$ & $6.288 \cdot 10^{-3}$ \\ 
2 & $16$ & $449{,}973$ & $\mathbf{3.346\cdot 10^{-2}}$ & $\mathbf{6.149 \cdot 10^{-3}}$ \\ 
3 & $32$ & $698{,}353$ & $3.350\cdot 10^{-2}$ & $6.279 \cdot 10^{-3}$ \\ 
4 & $64$ & $1{,}452{,}249$ & $3.363\cdot 10^{-2}$ & $6.271 \cdot 10^{-3}$ \\ \hline
\end{tabular}
\end{table}

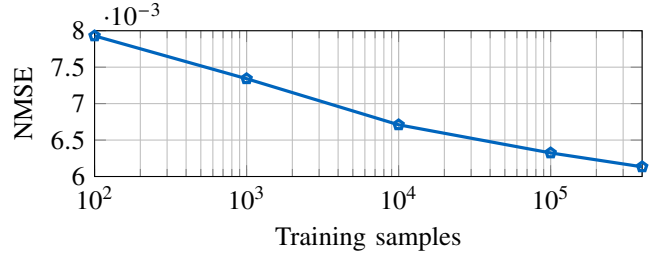
\begin{figure}[t]
    \begin{tikzpicture}
	\begin{axis}
		[   
		xlabel={Training samples},
		ylabel={NMSE},
		xmin=100, xmax=400000,
        ymin=6e-3, ymax=8e-3,
		xmode=log,
		ymajorgrids=true,
		xmajorgrids=true,
		grid style=solid,
		grid=both,
		legend pos=north east,
		width=\pltw,
		height=\plthsmall,
		legend style={font=\scriptsize},
		legend columns=1,
		legend style={at={(1,1)},anchor=north east}
		] 
  
		
		\addplot[VAEreal]
		table [ignore chars=", x=s, y=real, col sep=comma]{data/training_samples_20db.txt};
		
	\end{axis}
\end{tikzpicture}
    \vspace{-3mm}
    \caption{\ac{nmse} for different numbers of training samples at an SNR of 20\,dB for the measurement data channels.}
    \label{fig:training_samples}
\end{figure}

As pointed out in the previous subsection, the encoder input is mapped to \textsc{ch} \acp{cl}, which is multiplied by $1.75$ after each following \ac{cl}.
Since \textsc{ch} controls the number of parameters in the \ac{vae}, we want to investigate the relation between the number of parameters and the estimation performance by varying \textsc{ch}.
To this end, we display different model sizes and their performance at 10 and \SI{20}{dB} on the measurement test dataset in Table~\ref{tab:parameters}.
We observe that the second model achieves the lowest NMSE at both \ac{snr} values. 
Hence, making \textsc{ch} larger than $16$ does not improve the performance, so we utilize this value for \textsc{ch} in the remainder of this work.

To determine whether the full measurement training dataset achieves convergence of the model parameters, we illustrate the \ac{nmse} at \SI{20}{dB} on the measurement test dataset for the \ac{vae} estimator with different sizes of the training dataset, ranging from $100$ to $400{,}000$ in Fig.~\ref{fig:training_samples}.
As can be seen, the most significant performance gain is achieved until $100{,}000$ training samples are considered. 
When $400{,}000$ are considered, the \ac{nmse} still improves by a small value.
Fig.~\ref{fig:training_samples} highlights that a large training dataset is required for the considered model size to reach the full estimation potential.

\subsection{Estimation Results}
\label{subsec:estimation}

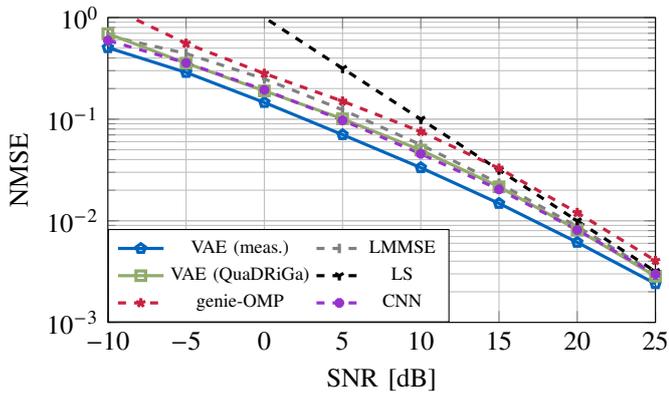
\begin{figure}[t]
    \begin{tikzpicture}
    	\begin{semilogyaxis}[
    		xlabel={SNR [dB]},
    		ylabel={NMSE},
    		xmin=-10, xmax=25,
            xtick={-10,-5,0,5,10,15,20,25,30},
    		ymajorgrids=true,
    		xmajorgrids=true,
    		grid=both,
    		grid style=solid,
    		legend pos=south west,
    		width=\pltw,
    		height=\plth,
    		legend style={font=\scriptsize},
    		legend columns=2,
    		ymin=1e-3, ymax=1,
		    legend style={at={(0,0)},anchor=south west}
    	] 
    	
    	
    	\addplot[VAEreal]
    		table [ignore chars=", x=snr, y=mse, col sep=comma]{data/results-mse-nokiameas-vae_real.txt};
    	    \addlegendentry{VAE (meas.)}
         
    	\addplot[globalcov]
    		table [ignore chars=", x=snr, y=mse, col sep=comma]{data/results-mse-nokiameas-global_cov.txt};
    	      \addlegendentry{\legglobalcov}
         
    	\addplot[mark options={solid}, color=TUMBeamerGreen, line width=\lineWidth, mark size=\markSize, mark=square]
    		table [ignore chars=", x=snr, y=mse, col sep=comma]{data/results-mse-nokiameas-vae_real_q.txt};
    	      \addlegendentry{VAE (\quadriga)}
    	
    	\addplot[LS]
    		table [ignore chars=", x=snr, y=mse, col sep=comma]{data/results-mse-nokiameas-ls.txt};
    	    \addlegendentry{\legLS}
    	
    	\addplot[genie OMP]
    		table [ignore chars=", x=snr, y=mse, col sep=comma]{data/results-mse-nokiameas-genie_omp.txt};
    	    \addlegendentry{\leggenieomp}
    	
    	
    	
    	\addplot[CNN]
    		table[x=SNR in dB, y=CNN, col sep=comma]
			{data/comparison_estimators_NMSE_Oct_URA_val_meas_64_comp.csv};
    		\addlegendentry{\legcnn}
    		
    	
    	\end{semilogyaxis}
\end{tikzpicture}
    \vspace{-3mm}
    \caption{Evaluation of the \ac{nmse} for the measurement test dataset. The proposed methods are displayed with solid linestyles.}
    \label{fig:nmse}
\end{figure}

We investigate the estimation performance in terms of \ac{nmse} on the measurement test dataset of the proposed \ac{vae}-based estimator in Fig.\ref{fig:nmse}. 
The word in brackets tells the dataset used for training the \ac{vae}, i.e., the blue curve belongs to the measurement training dataset and the green curve to the \quadriga training dataset.
The latter means that the \ac{vae} did not see any measurement dataset sample during its training.
As related baseline estimators, we display:
\begin{itemize}
    \item LMMSE: computes a global sample covariance matrix from the measurement training dataset and a \ac{lmmse} estimate as in~\cite{Baur2023}
    \item LS: a \ac{ls} estimate
    \item CNN: a \ac{cnn}-based estimator~\cite{Neumann2018}
    \item genie-OMP: the \ac{omp} algorithm that uses a four-times oversampled block-circulant matrix as a dictionary~\cite{Tropp2004}
\end{itemize}
In Fig.~\ref{fig:nmse}, it is visible that VAE (meas.) clearly outperforms all other methods at all \ac{snr} values.
Remarkably, VAE (\quadriga) performs almost identically as CNN, and both methods are the second best-performing estimators, followed by \ac{lmmse}.
A possible explanation for the fair performance of VAE (\quadriga) is that \quadriga generates propagation scenarios that resemble the measurement scenario.
Supposedly, the result is that both channel distributions are alike, which is beneficial for the online estimation phase.

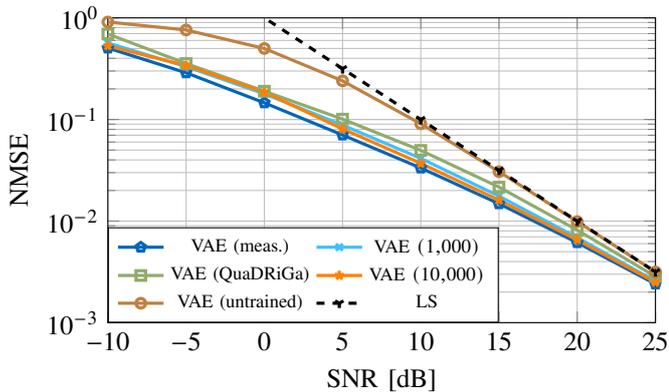
\begin{figure}[t]
    \begin{tikzpicture}
    	\begin{semilogyaxis}[
    		xlabel={SNR [dB]},
    		ylabel={NMSE},
    		xmin=-10, xmax=25,
            xtick={-10,-5,0,5,10,15,20,25,30},
    		ymajorgrids=true,
    		xmajorgrids=true,
    		grid=both,
    		grid style=solid,
    		legend pos=south west,
    		width=\pltw,
    		height=\plth,
    		legend style={font=\scriptsize},
    		legend columns=2,
    		ymin=1e-3, ymax=1,
		    legend style={at={(0,0)},anchor=south west}
    	] 
    	
    	\addplot[VAEreal]
    		table [ignore chars=", x=snr, y=mse, col sep=comma]{data/results-mse-nokiameas-vae_real.txt};
    	    \addlegendentry{VAE (meas.)}
    	
    	\addplot[mark options={solid}, color=TUMBeamerLightBlue, line width=\lineWidth, mark size=\markSize, mark=x]
    		table [ignore chars=", x=snr, y=mse, col sep=comma]{data/results-mse-nokiameas-vae_real_t_s_q_1000.txt};
    	    \addlegendentry{VAE $(1{,}000)$}
         
    	\addplot[mark options={solid}, color=TUMBeamerGreen, line width=\lineWidth, mark size=\markSize, mark=square]
    		table [ignore chars=", x=snr, y=mse, col sep=comma]{data/results-mse-nokiameas-vae_real_q.txt};
    	      \addlegendentry{VAE (\quadriga)}
    	
    	\addplot[mark options={solid}, color=TUMBeamerOrange, line width=\lineWidth, mark size=\markSize, mark=star]
    		table [ignore chars=", x=snr, y=mse, col sep=comma]{data/results-mse-nokiameas-vae_real_t_s_q_10000.txt};
    	    \addlegendentry{VAE $(10{,}000)$}
    		
    	\addplot[mark options={solid}, color=brown, line width=\lineWidth, mark size=\markSize, mark=o]
    		table [ignore chars=", x=snr, y=mse, col sep=comma]
    		{data/results-mse-nokiameas-vae_real_untrained.txt};
    	\addlegendentry{VAE (untrained)}
    	
    	\addplot[mark options={solid}, color=black, line width=\lineWidth, mark size=\markSize, dashed, mark=Mercedes star flipped]
    		table [ignore chars=", x=snr, y=mse, col sep=comma]{data/results-mse-nokiameas-ls.txt};
    	    \addlegendentry{\legLS}
    	
    	\end{semilogyaxis}
\end{tikzpicture}
    \vspace{-3mm}
    \caption{Evaluation of the \ac{nmse} for the measurement test dataset, which were pre-trained on \quadriga channel data.}
    \label{fig:pretraining}
\end{figure}

It is desirable to reduce the necessary training data of real-world measurements as much as possible.
One conceivable way is to pre-train the model on synthetic data and do the actual training on a smaller dataset.
We do this in Fig.~\ref{fig:pretraining} by pre-training \acp{vae} on the \quadriga training dataset.
The number in brackets is the number of training samples we use from the measurement data.
The blue and green curves represent the same models as in Fig.~\ref{fig:nmse}.
Fig.~\ref{fig:pretraining} shows that using a small number of measurement training data after pre-training with \quadriga data further improves the \ac{nmse}.
Compared with Fig.~\ref{fig:training_samples}, where no pre-training with \quadriga was performed, the relative performance gain with $1{,}000$ and $10{,}000$ measurement training samples is $6.3\,\%$ and $2.8\,\%$.
Thus, the gain with fewer samples is more noticeable.
We also plot a \ac{vae} with untrained weights only randomly initialized. 
This produces an identity matrix as estimated \ac{ccm} for every channel, which results in the \ac{nmse} 1/(\ac{snr}+1).
The explanation for this behavior is that we use an exponential function to enforce positive values in the vector $\vc_\vtheta(\vz)$, which receives only zeros as input due to the random weights.

\begin{figure}[t]
    \begin{tikzpicture}
    	\begin{semilogyaxis}[
    		xlabel={SNR [dB]},
    		ylabel={NMSE},
    		xmin=-10, xmax=25,
            xtick={-10,-5,0,5,10,15,20,25,30},
    		ymajorgrids=true,
    		xmajorgrids=true,
    		grid=both,
    		grid style=solid,
    		legend pos=south west,
    		width=\pltw,
    		height=\plth,
    		legend style={font=\scriptsize},
    		legend columns=2,
    		ymin=1e-3, ymax=1,
		    legend style={at={(0,0)},anchor=south west}
    	] 
    	
    	\addplot[VAEreal]
    		table [ignore chars=", x=snr, y=mse, col sep=comma]{data/results-mse-quadriga-vae_real_nokiameas.txt};
    	    \addlegendentry{VAE (meas.)}
         
    	\addplot[globalcov]
    		table [ignore chars=", x=snr, y=mse, col sep=comma]{data/results-mse-quadriga-global_cov.txt};
    	      \addlegendentry{\legglobalcov}
         
    	\addplot[mark options={solid}, color=TUMBeamerGreen, line width=\lineWidth, mark size=\markSize, mark=square]
    		table [ignore chars=", x=snr, y=mse, col sep=comma]{data/results-mse-quadriga-vae_real_quadriga.txt};
    	      \addlegendentry{VAE (\quadriga)}
    	
    	\addplot[LS]
    		table [ignore chars=", x=snr, y=mse, col sep=comma]{data/results-mse-quadriga-ls.txt};
    	    \addlegendentry{\legLS}
    	
    	\end{semilogyaxis}
\end{tikzpicture}
    \vspace{-3mm}
    \caption{Evaluation of the \ac{nmse} for the \quadriga test dataset.}
    \label{fig:eval_quadriga}
\end{figure}
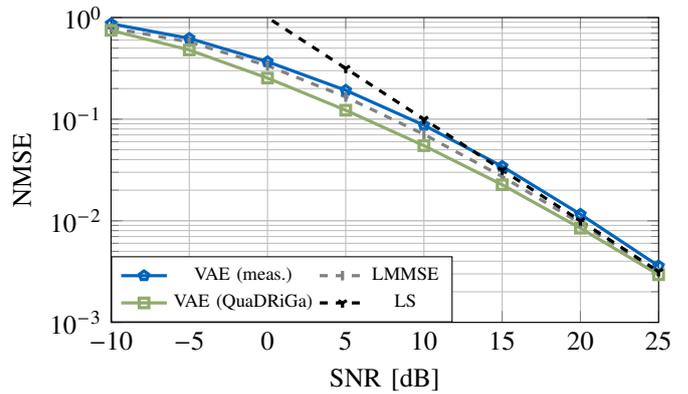

At last, we want to cross-check the previous analysis by evaluating a \ac{vae} on \quadriga data that is trained on the measurement data.
This investigation is presented in Fig.~\ref{fig:eval_quadriga}.
The plot shows a comparable performance gap between the two \acp{vae} as in Fig.~\ref{fig:nmse}, but in Fig.~\ref{fig:eval_quadriga} VAE (\quadriga) obviously (must) perform better.
Since VAE (meas.) performs worse than \ac{ls} in the high \ac{snr}, it can be concluded that it is not advisable to transfer a model from the measurement to the \quadriga data, which implies that the measurement data is more site-specific. 
However, the other direction appears reasonable based on the observations in Fig.~\ref{fig:nmse} and~\ref{fig:pretraining}.

\section{Conclusion}
\label{sec:conclusion}

This paper presents the evaluation of a \ac{vae}-based channel estimator on data stemming from real-world measurements.
The \ac{vae}-based estimator shows superior channel estimation performance compared to related state-of-the-art estimators. 
A downside of the \ac{vae}-based method is its requirement for an extensive training dataset to develop its full estimation potential.
Pre-training with synthetic channel data mitigates this necessity notably. 
However, since the \ac{vae} can be trained solely with noisy observations, the extensive training dataset requirement is not severe, as noisy observations can seamlessly be collected during regular \ac{bs} operation until the desired training dataset size is achieved.
Our future work includes the analysis of the \ac{vae}-based estimator in a real-world system.

\bibliographystyle{IEEEtran}
\bibliography{main}

\end{document}